\documentstyle[12pt]{article}
\def\be{\begin{equation}}
\def\ee{\end{equation}}
\def\bea{\begin{eqnarray}}
\def\eea{\end{eqnarray}}
\def\a{\alpha}

\def\e{\epsilon}
\def\d{\delta}

\author{Hans-J\"urgen Schmidt}

\title{The massive scalar field in a closed
 Friedmann universe model -- new rigorous results}

\date{}
\begin{document}
\maketitle

\centerline{Universit\"at Potsdam, Institut f\"ur Mathematik, Am
Neuen Palais 10} 
 \centerline{D-14469~Potsdam, Germany,  E-mail:
 hjschmi@rz.uni-potsdam.de}

\begin{abstract}
\noindent
For the minimally coupled scalar field
 in Einstein's theory of gravitation we 
look for the space of solutions within 
the class of closed Friedmann universe models.
 We prove $D \ge 1$, where $D$ is the dimension 
of the set of solutions which can be integrated up to 
 $t \to \infty$ 
 ($D > 0$ was conjectured by PAGE (1984)). 
We discuss concepts like ``the probability of the 
appearance of a sufficiently long inflationary phase" 
and argue that it is primarily a probability measure $\mu$
 in the space $V$ of solutions (and not in the space of 
initial conditions) which has to be  applied. 
$\mu$ is naturally defined for Bianchi-type I
 cosmological models because $V$  is a compact cube. 
The problems with the closed Friedmann model 
(which led to controversial claims in the literature) 
will be shown to originate from the fact that $V$ has a 
complicated non-compact non-Hausdorff Geroch topology: 
no natural definition of $\mu$ can  be  given.

\medskip

We conclude: the present state of our universe 
can be explained by models of the type discussed, 
but thereby the anthropic principle cannot be fully circumvented.

\bigskip
\noindent 
F\"ur das minimal gekoppelte Skalarfeld in 
Einsteinscher Gravitationstheorie betrachten wir 
den L\"osungsraum in der Klasse der geschlossenen 
Friedmannmodelle des Universums. Wir  
beweisen,  da\ss \ $D \ge 1$ gilt, wobei $D$ die Dimension 
der Menge derjenigen L\"osungen ist, 
die sich bis $t \to \infty$ integrieren lassen. 
($D > 0$ war von PAGE (1984) vermutet worden.) 
Wir diskutieren Begriffe wie ``die Wahrscheinlichkeit des 
Auftretens einer hinreichend langen inflation\"aren Phase''   
und argumentieren, da\ss \  man prim\"ar ein Wahrscheinlichkeitsma\ss \ 
$\mu$ im Raum der L\"osungen $V$ (und nicht im Raum der Anfangswerte) 
anwenden mu\ss. F\"ur kosmologische Modelle vom Bianchi-Typ I
 gibt es eine nat\"urliche Definition f\"ur $\mu$, da dort $V$ ein 
kompakter 
W\"urfel ist. Die Probleme mit dem geschlossenen Friedmannmodell 
(die zu Kontroversen in der Literatur f\"uhrten) ergeben sich aus der 
Tatsache, da\ss \  $V$ dort eine komplizierte Geroch-Topologie hat, 
die weder kompakt noch hausdorffsch ist, so da\ss \ sich keine 
nat\"urliche Definition f\"ur $\mu$ angeben l\"a\ss t.

\medskip

Wir schlie\ss en: Zwar l\"a\ss t sich der heutige Zustand 
unseres Universums durch Modelle des hier diskutierten 
Typs erkl\"aren, jedoch l\"a\ss t sich dabei das anthropische Prinzip 
nicht v\"ollig umgehen.
\end{abstract}


AAA subject classification: 161

\section{Introduction}

We consider a closed Friedmann cosmological model,
\be
ds^2 = 
g_{ij} dx^i dx^j =
dt^2 - a^2(t) [dr^2 + \sin^2 r(d \psi^2  + \sin^2 \psi d\chi^2)] \, ,
\ee
with the cosmic scale factor $a(t)$. We 
apply Einstein's General Relativity Theory 
 and take a minimally coupled scalar field $\phi$
without self-interaction 
as source, i.e., the Lagrangian is ($\hbar  = c =1$)
\be
{\cal L} = \frac{R}{16\pi G} + \frac{1}{2} g^{ij}
\nabla_i \nabla_j
- \frac{1}{2} m^2 \phi^2 \, .
\ee
$R$ is the scalar curvature, $G$ Newton's constant, and $m$ the mass 
of the scalar field. $\delta {\cal L}/\delta \phi=0$ yields
\be
     (m^2+ \Box) \phi = 0\,  ,   \qquad \Box \equiv  g^{ij}
\nabla_i \nabla_j
\ee
where $\Box$  is the covariant D'Alembertian. 
$$
\delta {\cal L}\sqrt{ - {\rm det} g_{kl}}
/\delta g^{ij}=0
$$
yields Einstein's equation
\be
R_{ij } - \frac{R}{2} g_{ij} = 8 \pi G T_{ij} \, , \quad
 T_{ij} \equiv \nabla_i \phi \nabla_j \phi
- \frac{1}{2} \left(\nabla^k \phi \nabla_k \phi - m^2 \phi^2
\right) \, , 
\ee
with Ricci tensor $R_{ij}$. It is the aim of the present paper to give 
some 
new rigorous results about the space $V$ of solutions 
of eqs. (3.4) with metric  (1)  (Scts. 2, 3, 4) and to discuss them in 
the context of cosmology (Sct. 5): the probability  of  the appearance 
of a sufficiently long inflationary phase and the anthropic principle.

\bigskip

SCHR\"ODINGER (1938) (here cited from SCHR\"ODINGER (1956)) 
already deals with the massive scalar field in the closed Friedmann 
universe, but there the back-reaction of the scalar field on 
the evolution of the cosmic scale factor had been neglected. 
35 years later the model enjoyed a renewed interest, especially as a 
semi-classical description of quantum effects, cf. e.g. 
FULLING (1973), FULLING and PARKER (1974), 
STAROBINSKY (1978), BARROW and MATZNER (1980), 
GOTTL\"OBER (1984a, b) and HAWKING and LUTTRELL (1984). One intriguing 
property - the possibility of a bounce, i.e., a positive local minimum of
 the cosmic scale factor - made it interesting in connection with a possible 
avoidance of a big bang singularity
 $ a(t) \to   0$. 
The existence of periodic solutions $(\phi(t), a(t))$  of 
eqs. (1, 3, 4)  became  clear in 1984  by  independent
work of  HAWKING  (1984a, b) and GOTTL\"OBER
 and SCHMIDT (1984). In PAGE (1984) it was conjectured 
(I see no chance to exclude that most 
of the sets  
$$
S_{n_1 \dots n_j}
$$
eq. (29) of that paper are empty) that besides 
the periodic solutions also a fractal set of Hausdorff dimension 
$D > 0$ of aperiodic perpetually bouncing universes exist. 
We shall prove this conjecture in a stronger version 
($D \ge 1$) in Sct. 3.

\bigskip

The existence of an inflationary phase 
(defined by $\vert dh/dt \vert \ll  h^2$, where $h = d (\ln a)/dt$   is the 
Hubble parameter) in the cosmic evolution is discussed 
in many papers to that model, cf e.g. 
BELINSKY et al. (1985), BELINSKY and 
KHALATNIKOV (1987),  PAGE (1987), 
GOTTL\"OBER and M\"ULLER (1987), and GOTTL\"OBER (1988).

\bigskip

Soon it became clear that the satisfactory results obtained for 
the spatially flat model --   the existence of a naturally defined 
measure in the space of solutions and with this measure the
very large probability to have sufficient inflation -- cannot 
be generalized to the closed model easily. We shall turn to that point in 
Sct. 5.

\section{Some closed-form approximations}

We consider the system (3, 4) with
 metric (1). Sometimes, one takes it as an additional 
assumption that $\phi$ depends on the coordinate $t$ only, 
but it holds (cf. TURNER 1983): the spatial homogeneity 
of metric (1) already implies this property.

\noindent
{\it Proof:}  For $i \ne j$ we have $R_{ij} =g_{ij} = 0$ and, therefore, 
$ \nabla_i \phi \nabla_j \phi= 0$: 
locally, $\phi$ depends 
on one coordinate $x^i$  only. Supposed $i  \ne  0$, 
then $g^{ij}T_{ij}    - 2T_{00} = m^2 \phi^2$.
      The l.h.s. depends on $t$ only. For $m \ne 0$ this is 
already a contradiction. For $m = 0$  we
have
$R_{ij} = \nabla_i \phi \nabla_j \phi$, 
which is a contradiction to the spatial isotropy of $R_{ij}$, q.e.d.

\bigskip

We always require $a(t) > 0$, for otherwise the 
metric (1) is degenerated (``big bang"). Eqs. (3, 4) reduce to
\be
m^2 \phi + d^2\phi/dt^2 + 3hd\phi/dt = 0
\ee
and
\be
     3\left( h^2 + a^{-2}\right) = 4\pi G[m^2\phi^2 + (d\phi/dt)^2] \,  . 
\ee
Eq. (6) is the 00 component of eq. (4), the other components are 
a consequence of it. 

\bigskip 

For the massless case
  $m= 0$  (i.e. stiff matter, pressure equals 
energy density), eqs. (5, 6) can be integrated. Let $\psi  = d\phi/dt$, 
then eq. (5) implies $\psi a^3 =$ const. Inserting 
this into eq. (6), we get 
$$
da/dt = \pm\sqrt{L^4/a^4 -1}
$$
 with a constant $L > 0$ and
$    0 < a \le L$. We get $a(t)$ via the inverted function up to a $ t$ 
translation
$$
    \pm  t = \frac{1}{2}L \arcsin (a/L)
          - \frac{a}{2} 
 \sqrt{1 - a^2/L^2} \, .
$$

\bigskip

\noindent 
This is conformally equivalent to $L = R^2$, cf. STAROBINKY 
and SCHMIDT (1987) and BARROW and SIROUSSE-ZIA (1989). 
For $a \ll L$ we have $a \sim  t^{1/3}$  and 
 $\phi = \phi_0 + \phi_1 \ln t $. 
  For $a = L$ 
we have a maximum of the function $a(t)$. $a(t) > 0$ is fulfilled for a $t$ 
interval of length $\Delta t = \pi L/2$ only, and there is no bounce.

\bigskip

In which range can one except the massless case to be a 
good approximation for the massive case? To this end we perform 
the following substitutions:
$$
   \tilde  t=t/\e\, , \quad \tilde a=a/\e \, , \quad
\tilde m=m \e \,  , \quad \tilde \phi =\phi \,  .
$$
They do not change the differential equations. Therefore, to get a 
solution $a(t)$ with a maximum 
$a_{\rm max} = \e \ll 1$ for $m$ fixed, we can transform 
to a solution with  $\tilde  a_{\rm max}= 1$, 
 $\tilde m = \e m \ll m$  and apply solutions  
$\tilde a \left( \tilde t \right)$
 with $\tilde m = 0$ as a good approximation. Then $a(t) > 0$  is 
fulfilled 
for a $t$ interval of length $\Delta t \approx \pi     a_{\rm max}/2$
 only. This is in quite 
good agreement with the estimate  in eq. (22) of PAGE (1984). 
But in the contrary to the massless  case it holds: to each $\e > 0$
 there 
exist bouncing solutions which possess a local maximum 
$a_{\rm max} = \e$.

\bigskip

Let henceforth be $m > 0$. Then we use   
 $ \sqrt{4\pi G/3}
\phi$
instead of $\phi$ as scalar 
field and take units such $m = 1$. With a dot denoting 
$d/dt$ we finally get from eqs. (5, 6)
\be
\phi + \ddot \phi  + 3h\dot \phi  = 0 \, ,
\quad   h^2 + a^{-2} = \phi^2 + \dot \phi^2 \, .
\ee
$\phi \to - \phi$ is a 
$Z_2$-gauge transformation ($Z_2$ is the two-point group). 
Derivating 
eq. (7) we 
can express $\phi$ and 
$ \dot \phi$
as 
follows 
\be
\phi = \pm \sqrt{2+2\dot a^2+   a \ddot a} / \sqrt{3a^2}
\, , \quad 
\dot \phi = \pm \sqrt{1+\dot a^2 -  a \ddot a} / \sqrt{3a^2} \, .
\ee
Inserting
(8)  into eq. (7) we get
\be
a^2 \frac{d^3 a}{dt^3} = 4 \dot a(1+\dot a^2) - 3 a \dot a \ddot a
 \pm 2a  \sqrt{2+2\dot a^2+   a \ddot a}
\sqrt{1+\dot a^2 -  a \ddot a} \, .
\ee
On the r.h.s. we have $``+"$
 if   $\phi \dot \phi  > 0$ and $``-"$ otherwise. At 
all points $t$, where one of the roots becomes zero,
 $``+"$ and $``-"$ 
 have  to be interchanged.

\bigskip

To get the temporal behaviour for very large 
values $a \gg 1$ but small values $\vert h \vert$,
 we make the ansatz
\be
     a(t) = 1/\e + \e A(t)\, , \quad  \e>0\, , \quad  \e \approx 0\, .  
\ee
In lowest order of $\e$ we get from (9)
\be
\frac{d^3 A}{dt^3} = \pm 2 \sqrt{2+\ddot A}\sqrt{1-\ddot A} \, .
\ee
An additive constant to $A(t)$ can be absorbed by a redefinition of $\e$, 
eq. (10), so we require $A(0)=0$. After a suitable translation of $t$, 
each solution of eq. (11) can be represented as
\be
     A(t) = \a t - t^2/4 + \frac{3}{8}
 \sin(2t) \, , \quad
     \vert \a \vert \le \pi/4 \, .
\ee
$\a = \pi/4$ and $\a = - \pi/4$
 represent the same solution, so 
we have a $S^1$ space of solutions ($S^n$ is 
the $n$-dimensional sphere). $a(t) > 0$ is fulfilled for 
$A(t) > - 1/\e^2$,  i.e. $\vert t\vert 
<  2/\e $ only. In dependence of the value $\a$, $A(t)$ has one or two 
maxima and, accordingly, null or one minimum. The corresponding 
intervals for $\a$ meet at two points, $\a \approx  0$ and 
$\vert \a \vert \approx  \pi/4$, where one 
maximum and one horizontal turning point
$$
\dot a = \ddot a =  0, \ \frac{d^3 a}{dt^3}  \ne 0
$$
exist.

\section{The qualitative behaviour}

Eq. (7) is a regular system: at $t =  0$ we prescribe 
$a_0$, sgn $\dot a_0$, $\phi_0$, $\dot \phi_0$ 
fulfilling
$$
a_0^2 \left (\phi_0^2 + \dot \phi_0^2 \right) \le 1
$$
 and then 
all higher derivatives can be obtained by differentiating:
$$
\vert \dot a \vert = \sqrt{a^2(\phi^2 + \dot \phi^2) -1} \, , 
\ddot \phi = - 3 \dot a \dot \phi/a - \phi \, , 
$$
the r.h.s. being smooth functions. 
It follows
\be
d/dt \left(h^2 + a^{-2}\right) = -6h(d\phi/dt)^2 \, .
\ee

\subsection{Existence of a maximum}

The existence of a local maximum for each
 solution $a(t)$  already follows from the ``closed universe 
recollapse conjecture", but we shall prove it for our model as 
follows: if we start integrating with $ h \ge 0 $
 (otherwise, $t \to   -t$ serves 
to reach that), then $h^2 + a^{-2}$  is a monotoneously decreasing 
function as long as $h \ge 0$ holds ($h \dot \phi  = 0$
 holds at isolated points $t$ only), 
cf. eq. (13). We want to show that after a finite time, $h$ changes 
its sign giving rise to a local maximum of $a(t)$. If this is not the 
case after a short time, then we have after a long time 
$h \ll 1$, $a \gg 1$, 
and eqs. (10, 12) become a good approximation to the exact solution. 
The approximate solution (10, 12) has already shown to possess 
a local quadratic maximum and this property is a stable one 
within $C^2$ perturbations. So the exact solution has a maximum, too, 
q.e.d.

\subsection{The space of solutions}

We denote the space of solutions for (1,7) by $V$ 
and endow it with Geroch's topology  (GEROCH 1969); 
see Sct. 4.3. for further details. 
Applying the result of Sct. 3.1., $V$ can be 
constructed as follows: the set of solutions will 
not be diminished if we start integrating with 
$\dot a_0 = 0$. We prescribe
\be
f= a_0^{-1} \quad {\rm  and } \quad 
 g = (1 - a_0 \ddot a_0) \cdot {\rm sgn} 
(\phi_0 \dot \phi_0) \, ,
\ee
then all other values are fixed. $f$ and $g$ are restricted 
by $f> 0$; $\vert g \vert \le 3$, where $g = 3$ and $ g = -3$ describe 
the same 
solution. Therefore
\be
V    = ({\cal R} \times S^1)/Q \, ,
\ee
where ${\cal R} $ denotes the real line 
(as topological space) and $Q$ is an equivalence  relation 
defined as follows:

\bigskip

Some solutions $a(t)$  have more  than one (but 
at most countably many) extremal points, but each
 extremum defines one point in $  {\cal R} \times S^1  $; these are 
just 
the points being $Q$-equivalent.

\bigskip

Considering eqs. (7, 14) in more details, we get the 
following: for $\vert g \vert  < 1$ we have a minimum and for 
$\vert g \vert > 1$ 
 a maximum for $a(t)$. 
$$ \frac{
d^3 a_0}{dt^3} >
 0
$$
 holds  for $0 < g < 3$. $a(t)$ is an even 
function for $g = 0$ and $g =\pm 3$ only. For $g = 0$ it is 
 a symmetric 
minimum of $a(t)$  and $\phi$ is even, too. For $g = \pm 3$
 it is a symmetric
maximum of $a(t)$  and $\phi$ is odd. For $\vert g \vert  = 1$
 one has a 
horizontal turning point for $a(t)$. Time reversal $t \to   -t$
 leads to $g \to  -g$.

\subsection{From one extremum to the next}

To prove our result, $D \ge 1$,  where $D$ is the dimension
 of the set of solutions which can be integrated up to $t \to \infty$, 
we need a better knowledge of the equivalence relation $Q$. 
To this end we define a map 
$$
p: {\cal R} \times S^1   \to {\cal R} \times S^1
$$
 as follows:

\bigskip

For $x = (f, \, g)$ we start integrating at $t = 0$.
 Let 
$$
t_1 = {\rm min} \{t \vert t > 0, \,  \dot a(t) = 0\} \, .
$$ 
For $t_1 < \infty$  we define 
$$
p(x) = \bar x = (\bar f, \, \bar g) 
 = \left( f(t_1), \, g(t_1) \right) 
$$
and otherwise $p$ is not defined. (This notation 
applies only to this Sct. 3.)

\bigskip

In words: $p$ maps the initial conditions from one 
extremal point of $a(t)$ to the next one. It holds: $p$ is 
injective and $xQy$ if and only if there exists an 
integer $m$ such that $p^m(x) = y$.

\bigskip

Let $V_m \subset {\cal R} \times S^1$ be that
 subspace for which $p^m$ is defined. 
$$
\{(f, \, g) \vert - 1 <g \le 1 \} \subset  V_1
$$
means:  a minimum is always followed  by 
a maximum; $\vert g \vert  < 1$
 implies $\vert \bar g \vert \ge 1$.
 $V_2 \ne \emptyset$  follows from the 
end of Sct. 2.  For $x = (f, \,  g)$ we define 
 $-x = (f, \, - g)$. 
With this 
notation it holds (see the end of  Sct. 3.2) 
$$
p^m(V_m) = V_{-m} = - V_m \, ,
$$
 and for $x \in V$  we  have
\be
    p^m \left( -p^m(x)\right) = -x
 \quad {\rm
 and} \quad 
 p^{-m}\left( V_m \cap V_{-m} \right) 
 = V_{2m} \, .
\ee
A horizontal turning point can be continuously deformed
 to a pair 
of extrema; such points give rise to discontinuities of the function 
$p$, 
but for a suitably defined nonconstant integer power $m$
 the function $p^m$ is 
a continuous one. If $x \in V_1 \backslash {\rm int}(V_1)$ 
 (int denotes the topological interior), 
then $a(t)$ possesses a horizontal turning point. To elucidate
 the contents 
of these sentences we give an example:

\bigskip

For very small values $f$  and $\bar f$ we may apply eqs. (10, 12) to 
calculate 
the function $p^m$.
 In the approximation used, no more  than 3 extremal 
points appear, so we have to consider 
$$
 p^m \, , \quad m \in \{-2, \,  - 1, \,  0, \, 1, \,  2 \}
$$ 
only. $f$ and $\bar f$  approximately coincide, so  we 
concentrate on the function $\bar g(g)$, 
see Fig. 1. The necessary power $m$ is sketched at the curve.
 To come 
from $m$ to $-m$, the curve has to he reflected at the line 
$\bar g = g$. Time reversal 
can be achieved, if it is reflected at $\bar g = -g$. $V_1$
 is the interval $
-1 <g \le  2.8853$. The jump discontinuity of $p$ at $g = 1$ and the 
boundary value $g =2.8853$ \dots  are both connected with  horizontal 
turning points $g, \, \bar g = \pm 1$. The shape of the 
function $\bar g(g)$ in Fig. 1 for 
$m = 1$ can be obtained by calculating  the extrema of $A(t)$
 (eq. (12)) 
in dependence of $\a$
 and then applying eq. (14). For $m \ne 1$ one applies eq. (16).

\bigskip

Fig. 1. Poincare-return map for 
the closed Friedmann model, 
explanation see Sct. 3.3, and $m =n$. 

\bigskip

\subsection{The periodic solutions}

The periodic solutions $a(t)$ are characterized by 
the fixed points of some $p^m$, $ m \ge 2$. ($p$  itself has no 
fixed points). The existence  of them can be proved as follows: 
we start integrating at $t = 0$ with  a symmetric minimum 
($g = 0$) of $a(t)$  and count the  number $m$ of zeros of $\phi(t)$
 in the 
time interval $0 <t < t_2$, where $a(t_2)$  is the first 1ocal 
maximum of $ a(t)$. $m$ depends on $f = 1/a(0)$  and has jump 
discontinuities only at points where $\phi(t_2) = 0$.  For initial 
values $f = f_k$, where $m$ jumps from $k$  to $k + 1$, $a(t)$
 is symmetric 
about $t = t_2$. But a function, which is symmetric about two 
different points, is a periodic one.  We call them periodic solutions 
of the first type, they are fixed points of $p^2$.

\bigskip

PAGE (1984) gives the numerically obtained 
result: $f_1 = 1/a_0$ with $a_0=0.76207$\dots  That 
$f_k$  exists also  for 
very large values $k$ can be seen as follows: take solution (10, 12) 
with
 a 
 very small value $\e$
 and insert it into eq. (8); we roughly get
 $ m = 1/\e \pi$.

\bigskip

Periodic solutions of the second type, fixed points 
of $p^4$, which are not fixed points of $p^2$, can be constructed
 as follows:  we 
start with $g = 0$ but $f \ne f_k$  and count the number 
$n$ of  zeros of $\dot \phi(t)$ in the 
interval $0 < t < t_3$, where $ a(t_3) > 0$
 is next local minimum of $a(t)$. 
(For $f \approx  f_k$, $t_3$ is defined by continuity reasons).

\bigskip

For $f = f^l$, $n$ jumps from $l$ to $l + 1$ and we have a periodic 
solution. Numerical evaluations yield $f^1 = 1/a_0$,
 $a_0 = 0.74720$\dots The 
existence of $f^l$ for very large values $l$
 is again ensured by solution (10, 12).

\subsection{The aperiodic  perpetually oscillating solutions}

Now we look at the solutions in the neighbourhood of  the 
periodic ones. Let $x_0 = (f_k, \, 0) \in V_2$ be one of the fixed 
points of $p^2$; $a_0(t)$, the corresponding periodic solution of the 
first 
type has
 no horizontal turning points. Therefore, $p$ is smooth at $x_0$. 
Let us denote the circle with boundary of radius $\e$
 around $x_0$ by $K(\e)$. 
By continuity reasons there exists an $\e > 0$ such that 
$K(\e) \subset {\rm  
int} ( V_2 \cap V_{-2})$ and the first two extrema of the functions  
$a(t)$ corresponding to points of $K(\e)$  are either maxima or 
minima. 
Let   $R_0 = K(\e)$  and for $n \ge 1$
$$
     R_n: = R_0 \cap p^2 \left( R_{n-1}\right) \, ; 
 \quad       R_{-n}: = R_0 \cap p^{-2} \left( R_{1-n}\right)
\, . 
$$
By assumption, $p^{\pm 2}$
 is defined in $R_0$, so $R_m$  is a well-defined compact 
set with $x_0 \in {\rm int} \, R_m$ for each integer $m$. It
 holds
$$
 R_{n+1} \subset R_n \, , \quad R_{-n} = - R_n \, 
$$
and 
$$
R_n = \{ x \vert p^{2m}(x) \in R_0 \ {\rm for} \quad 
m = 0, \dots n \} \, , 
$$
$$
R_\infty := \bigcap_{n=0}^\infty R_n
$$
is 
a compact set with $x_0 \in  R_\infty$.

\bigskip

For each $x \in  R_\infty$,  the 
corresponding solution $a(t)$ can be integrated up to $t \to \infty$. 
The last statement follows from the fact
that the time from one extremum to the next is bounded from 
below by a positive number within the compact set $R_0$, i.e., an 
infinite number of extrema can be covered only by an infinite 
amount of time. Analogous statements hold for   
$R_{- \infty} =  - R_\infty$ and  $t \to - \infty$. 

\bigskip

Let us fix an integer $m \ge 1$. 
We start integrating at $x_\d := (f_k, \, \d ) \in R_0$, $0 < \d \le \e$.
$$
\d(m) := {\rm max} \{ \d \vert \d \le \e, \ p^{2k}(x_\d) \in R_0 \ 
{\rm for } \
 k=0, \dots m \}
$$
exists because of compactness, i.e., 
$x_{\d(m)} \in 
 R_m$   and there is an 
integer $k(m) \le  m$  such that
$$
 y(m) : = p^{2k(m)} \left(x_{\d(m)} \right) \in \d R_0   \, , 
$$
 $ \d R_0 $ 
being the boundary of $R_0$.
$$ 
    j(m) := 1\,  , \quad   {\rm   if } \quad  k(m) < m/2 \, ,
 \quad {\rm  otherwise} = 0 \, .
$$
The sequence 
$$
\left(y(m), \,  j(m) \right) \subset  S^1 \times  Z_2
$$
 possesses a 
converging subsequence with 
$$
y(\infty):= \lim_{i \to \infty} y(m_i)\, , \quad 
j(\infty):= \lim_{i \to \infty} j(m_i) \, . 
$$

\bigskip

Now let us start integrating  from $y(\infty)$ forward in 
time for $j(\infty) = 1$  and backwards in time for 
$j(\infty) =  0$. We consider 
only the case $j(\infty)=  1$, the other case will be solved by 
$t \to  - t$. For 
each $i$
 with $m_i  > 2m$  we have $y(m_i) \in R_m$,
 therefore, $y(\infty) \in  R_m$  
for all $m$, i.e., $ y(\infty) \in
 R_\infty$. $y(\infty) \in  \d R_0$  and for all 
 $n \ge 1$, $ p^{2n} \left( y(\infty) \right) \in R_0$.

\bigskip

By continuously diminishing $\e$ we get a one-parameter set 
of solutions $a_\e(t)$, which can be integrated up to $t \to \infty$.
 (To this end 
remember that one solution $a(t)$  is represented by at most 
countably many points of $R_0$). 
Supposed $a_\e(t)$  is a periodic function. 
By construction this solution has a symmetric minimum and there exist 
only countably many such solutions. Let $M = \{a(t) \vert a(0)$  is a 
minimum 
parametrized by a point  of $R_0$, $a(t_1)$  is the next minimum, and 
 $0 \le t \le t_1 \}$  then $M=[ a_{\rm min}, \, a_{\rm max}] $
with 
$ a_{\rm min} > 0$, $ a_{\rm max} < \infty $
and for each $\e$ and each $t \ge 0$ it  holds 
$ a_{\rm min} \le a_\e(t) \le  a_{\rm max}  $. 

\bigskip

So we have proven: in each neighbourhood of the periodic solutions of 
the first type there exists a set of Hausdorff  dimension $D \ge 1$  
of 
uniformly bounded aperiodic perpetually oscillating solutions, which 
can 
be integrated up to $ t \to \infty$.

\section{Problems with die probability measure}

To give concepts like ``the probability $p$ of the
 appearance of  a sufficiently long inflationary phase''    a 
concrete meaning, we have to define a probability measure $\mu$  in 
the space $V$ of solutions. Let us suppose we can find a hypersurface 
 $H$  in the space $G$ of initial conditions such that each solution 
is characterized by exactly one point of $H$. Then $V$ and $H$ are 
homeomorphic and we need not to make a distinction between them. 
Let us further suppose that $H \subset G$ is defined by a suitably 
chosen physical quantity $\psi$ to take the Planckian value. Then 
we are justified to call $H$ the quantum boundary. By construction, 
$H$ divides $G$ into two connected components; $\psi \le \psi_{\rm Pl}$ 
defines the classical region. All classical trajectories start their 
evolution 
at $H$ and remain in the classical region forever. 

\bigskip

Let us remember the situation for the 
spatially flat Friedmann model (BELINSKY et al. (1985), 
 Sct. 4.1.) and for the Bianchi type I model (LUKASH  and 
SCHMIDT (1988), Sct. 4.2) before  we discuss the closed 
Friedmann model in Sct. 4.3.

\subsection{The spatially flat Friedmann model}

For this case, eq. (13) reduces  to 
$h \dot h = -3h \dot \phi^2$,
 i.e., each solution crosses the surface $h = h_{\rm Pl}$  
exactly once, the only exception is the flat Minkowski space-time 
 $h \equiv  0$. 
And the corresponding physical quantity
\be
\psi = h^2 = \phi^2 + \dot \phi^2
\ee
is the energy density. The space of non-flat spatially flat Friedmann 
models is topologically $S^1$,
 and equipartition of initial conditions 
gives a natural probability measure there. With this definition it 
turned 
out that for $ m \ll m_{\rm Pl}$  it holds
$$
p \approx   1 - 8m/ m_{\rm Pl} \, ,
$$
 cf. e.g. M\"ULLER and SCHMIDT (1989), i.e., 
inflation becomes quite probable. If, 
on the other hand, equipartition is taken at some 
$h_0 \ll h_{\rm Pl} m /m_{\rm Pl}$
 then inflation is quite improbable.

\bigskip

The total space $V$  of  solutions  has Geroch 
topology    $ \alpha S^1$,
 i.e., $V = S^1 \cup \{ \alpha \}$,
 and the space itself is the only 
neighbourhood  around the added point $\alpha$
 (which corresponds to $h \equiv 0$), 
because each solution is asymptotically flat  for $t \to \infty$.

\subsection{The Bianchi-type I  model}

With the metric
$$
     ds^2 = dt^2
 - e^{2 \a} [e^{2(s+ \sqrt 3 r)} dx^2
 + e^{2(s - \sqrt 3 r)} dy^2 + e^{-4s}   dz^2 ], \quad
    h = \dot \a
$$
the analogue to eq. (17) is
\be
\psi
 = h^2 = \phi^2 + \dot \phi^2 + \dot r^2 + \dot s^2 
\ee
and $h =   h_{\rm Pl}$
 defines a sphere $S^3$ in eq. (18). Here
 all solutions cross this sphere exactly once, even the flat Minkowski 
spacetime: it is represented as (0 0 l)-Kasner solution $\a$,
 so the space of solutions is $V = S^3/Q$, where $Q$  is a 
12-fold cover of $S^3$ composed 
of the $Z_2$-gauge transformation 
 $\phi \to - \phi $
  and of the six permutations of 
the spatial axes.

\bigskip

Eq. (18) induces a natural probability measure on the space 
 $\psi = \psi_{\rm Pl}$ 
by equipartition,  and the equivalence relation $Q$ does not 
essentially 
influence this. As in Sct. 4.1., $\a \in V$  has only one 
neighbourhood: 
$V$ itself. Up to this exception, $V$ is  topologically a 
3-dimensional 
cube with boundary. One diagonal line through it represents the 
Kasner solution $\phi \equiv 0$.
The boundary $\delta V$ of $V$  has topology $S^2$  and 
represents the axially symmetric solutions and one great
 circle of it the isotropic 
ones. (As usual, the solutions
 with higher symmetry form the 
boundary of  the space of solutions.) $p$ turns out 
to be the same as in Sct. 4.1 
and how $\psi$ (eq. (18)) can be invariantly defined, is discussed 
in LUKASH and SCHMIDT (1988).

\subsection{The closed  Friedmann  model}

Now  we  come to the analogous questions concerning
 the closed Friedmann model.  Before defining a  measure, 
one should  
have a topology  in a set. I  feel it should be a variant of GEROCH (1969). 
The Geroch topology  (cf. SCHMIDT (1987), especially  footnote 
22) 
 is defined as follows: let $x_i = (a_i(t), \, \phi_i(t))$
  be a sequence 
of  solutions and  $x = (a(t), \, \phi(t))$  a further solution. 
 Then $x_i \to  x$ in 
Geroch's topology, if there exist suitable gauge and 
coordinate transformations after which $a_i(t) \to a(t)$ 
 and $\phi_i(t) \to \phi(t)$  
converge uniformly 
together with all their derivatives in the interval 
 $t \in  [- \epsilon, \, \epsilon ]$ 
for some $\epsilon > 0$. Because
 of the validity of the field 
equations, ``with all their derivatives" may be substituted 
by ``with their first derivatives".

\bigskip

With this definition one gets just 
the same space $V$ as in Sct. 3.2., eq. (15). The existence of aperiodic 
perpetually oscillating  solutions (which go right across the region
 of  astrophysical interest Sct. 3.5) shows that 
for a subset of dimension $D \ge 1$ of $R \times S^1$, $Q$
 identifies 
 countably many points. All these points lie in a compact 
neighbourhood of the corresponding periodic  point
 $ (f_k, \,  0)$ 
 and possess therefore (at least) one accumulation point $z$. 
At these points $z$, $V$  has a highly non-Euclidean  topology. 
Further, $V$  has a non-compact non-Hausdorff topology. 
So there is no chance to define a probability measure  in a natural 
way and no possibility to define a continuous hypersurface in the 
space of initial conditions which each trajectory crosses exactly once.
     
\section{Discussion}

Supposed, we had obtained a result of the 
type: ``Each solution $a(t)$  has at least one but at most seven 
local maxima." Then  one could define  --- up to a factor $7 =O(1)$
 ---  a 
probability measure. So it is just the existence of the 
perpetually bouncing  aperiodic solutions which gives the problems. 
We conclude: it is not a lack of mathematical knowledge but 
an inherent  property of the closed Friedmann model which hinders 
to generalize the convincing results obtained for the spatially 
flat  model. So it is no wonder that different trials led 
to controversial 
results, cf. BELINSKY et al. (1985, 1987) and PAGE 
(1987).  One of these results reads ``inflationary 
and noninflationary 
solutions have both infinite measure", hence, nothing is clear.  

\bigskip

By the way, PAGE (1987) claimed to have received 
different results for the massive scalar field in Einstein's 
theory  on the 
one hand and for $R + \epsilon R^2$  gravity on the other hand, 
e.g. the existence of singularity-free solutions for 
the open Friedmann  model only for the $R + \epsilon R^2$
 model. But all 
these singularity-free solutions cross the critical value 
 $R = R_c  = - 1 / 2 \epsilon$  
of the  curvature scalar which is known to be 
unstable: arbitrarily small anisotropies
 in the initial conditions will lead to
$$
C_{ijkl} C^{ijkl} \to \infty 
 \quad {\rm 
as } \quad  R \to R_c
$$
and, if we accept this point to be a singularity, 
too, then the results 
for both theories will again become the same.

\bigskip

Let us now discuss the results of STAROBINSKY (1978) 
and BARROW and MATZNER (1980) concerning the 
probability of a  bounce. They have obtained a very low 
probability 
to get a bounce, but they used equipartition at 
some $h_0 \ll h_{\rm Pl} m/ m_{\rm Pl}$. 
As seen in Sct. 4.1. for the spatially flat Friedmann model 
concerning inflation, this low probability does not hinder to 
get a considerable large probability if equipartition is applied at 
 $h = h_{\rm Pl}$.
 We conclude, the probability of bouncing solutions is not 
a well-defined concept up to now. Well-defined is, on the other 
hand, some type of conditional probability. If we suppose that 
some fixed value $a$, say $10^{28}$ cm or so, and there some fixed value
 $ h$, 
say 50 km/sec $\cdot$ Mpc or so, appear within the cosmic evolution, then 
the remaining degree of freedom is just the phase of
 the scalar field, which 
 is the compact set $S^1$  as configuration space. (But this solves 
not all problems, because the perpetually bouncing solutions discussed 
in Sct. 3.5. cross this range of astrophysical interest infinitely 
often.) Calculating conditional probabilities instead of absolute 
probabilities, and, if this condition is related to our own 
human existence, then we have already applied the anthropic principle. 
(Cf.  similar opinions in SINGH and PADMANABHAN (1988) concerning 
the so far proposed explanations 
of the smallness of the cosmological constant.)

\bigskip

The massive scalar field in  a closed Friedmann 
model with Einstein's  theory of gravity cannot explain 
the long inflationary stage of cosmic evolution as an absolute 
probable event and so some type of an anthropic principle has to 
be applied. See BARROW and TIPLER (1988) to this theme.

\bigskip

We have discussed the solutions for the minimally 
coupled scalar field, but many results for the conformally 
coupled one are similar (see e.g. TURNER and WIDROW 1988); 
this is explained by the existence of a conformal transformation 
relating between them (see SCHMIDT 1988).

\bigskip

The problems in the case of defining a probability 
measure in the set of (not necessarily spatially flat) Friedmann models 
are also discussed in MADSEN  and ELLIS (1988). 
They conclude that e.g. inflation need not to solve 
the flatness problem. (The Gibbons-Hawking-Stewart approach 
gives approximately the same probability measure as the equipartition of 
initial conditions used here.)

\noindent
{\it Acknowledgement.} I
 thank Drs. U. KASPER and V. M\"ULLER for critically reading the 
manuscript.

\section*{References}

\noindent
BARROW, J., SIROUSSE-ZIA, H.: 1989, Phys. Rev. D 39, 2187.

\noindent
BARROW, J., MATZNER, R.: 1980, Phys. Rev. D 21, 336.

\noindent
BARROW, J., TIPLER, F.: The anthropic cosmological principle, 
Cambridge University Press 1988.

\noindent
BELINSKY, V., GRISHCHUK, L., ZELDOVICH,  
Y., KHALATNIKOV, I.: l985, Zh. Eksp. Teor. Fiz.
89, 346.

\noindent
BELINSKY, V., KHALATNIKOV, I.: 
1987, Zh. Eksp. Teor. Fiz. 93, 784.

\noindent
FULLING, S.: 1973, Phys. Rev. D 7, 2850.

\noindent
FULLING, S., PARKER, L.: 1974, Ann. Phys. NY 87, 176.

\noindent
GEROCH, R.: 1969, Commun. Math. Phys.  13, 180.

\noindent
GOTTL\"OBER, S.: 1984a, Ann. Phys. Leipz. 41, 45.

\noindent
GOTTL\"OBER, S.: 1984b, Astron. Nachr. 305, 1.

\noindent
GOTTL\"OBER, S.: 
1988, in ``Gravitation und Kosmos" (Ed.: R. WAHSNER), Akad.-Verl.
Berlin, p. 50.

\noindent
GOTTL\"OBER, S., M\"ULLER, V.: 1987, 
Class. Quant. Grav. 4, 1427.

\noindent
GOTTL\"OBER, S.,
 SCHMIDT, H.-J.: 1984, Potsdamer Forschungen B43, 117.

\noindent
HAWKING, S.: 1984a, Nucl. Phys. B239, 257.

\noindent
HAWKING, S.: 1984b, ``Relativity, Groups, and Topology II" (Eds.:
B.   DE WITT,  R. STORA) North Holland
 PC Amsterdam.

\noindent 
 HAWKING, S., LUTTRELL,
J.: 1984, Nucl. Phys. B247, 250.

\noindent
LUKASH, V., SCHMIDT, H.-J.: 1988, Astron. Nachr. 309, 25.

\noindent
MADSEN, M., 
 ELLIS, G.: 1988, Mon. Not. R. Astron. Soc. 234, 67.

\noindent
M\"ULLER, V., SCHMIDT, H.-J.: 1989, Gen. Relat. Grav. 21, 489.

\noindent
PAGE, D.: 1984, Class. Quant. Grav. 1, 417.

\noindent
PAGE, D.: 1987, Phys. Rev. D 36, 1607.

\noindent
SCHMIDT, H.-J.: 1988, Phys. Lett. B214, 519.

\noindent
SCHMIDT, H.-J.: 1987, J. Math. Phys. 28, 1928.

\noindent
SCHR\"ODINGER, E.: 1938, Comment. Vatican. Acad. 2, 321.

\noindent
SCHR\"ODINGER, E.: 1956, 
Expanding Universes, Cambridge Univ. Press.

\noindent
SINGH, T.,
  PADMANABHAN, T.: 1988, Int. J. Mod. Phys. A3, 1593.

\noindent
STAROBINSKY, A.: 1978, Pisma v Astron. J. 4, 155.

\noindent
STAROBINSKY, A. A., SCHMIDT, H.-J.: 1987, Class. Quant. Grav. 4, 695.

\noindent
TURNER, M.: 1983, Phys. Rev. D 28, 1243.

\noindent
TURNER, M., WIDROW, L.: 1988, Phys. Rev. D 37, 3428.

\bigskip

\noindent 
{\it Note added in proof:} 

\noindent 
The Wheeler-de Witt equation for the 
massive scalar field in a closed Friedmann universe 
model is recently discussed by  E. CALZETTA
 (Class. Quant. Grav. {\bf 6} (1989) L227). 
Possibly, this approach is the route 
out of the problems mentioned here.

\bigskip

Received 1989 July 24

\bigskip

\noindent 
  In this reprint done with the kind permission of the 
copyright owner 
we removed only obvious misprints of the original, which
was published in Astronomische Nachrichten:   
 Astron. Nachr. {\bf 311} (1990) Nr. 2, pages 99 - 105.
 (The figure contained in the original paper is not reproduced here.)

\bigskip

\medskip
\noindent 
  Authors's address  that time: 

\noindent
H.-J. Schmidt,  
Zentralinstitut f\"ur  Astrophysik der AdW der DDR, 
1591 Potsdam, R.-Luxemburg-Str. 17a

\end{document}